\documentclass[twocolumn]{openjournal}

\usepackage{bm}
\usepackage{xspace}
\usepackage{xcolor}
\usepackage{amsmath}
\usepackage{graphicx}
\usepackage[hidelinks]{hyperref}
\hypersetup{breaklinks=true,colorlinks=true,linkcolor=blue,urlcolor=blue,citecolor=blue}
\usepackage[caption=false]{subfig}
\usepackage{booktabs}
\usepackage{orcidlink}

\usepackage{savesym}
\savesymbol{tablenum}
\usepackage{siunitx}
\restoresymbol{SIX}{tablenum}

\usepackage{tabularx}

\usepackage{lineno}

\usepackage{ragged2e}

\usepackage[capitalise]{cleveref}

\creflabelformat{equation}{#2\textup{#1}#3}
\newcommand{\Alens}{\ensuremath{A_{\rm LT}}}
\newcommand*{\rgw}{\ensuremath{r}}
\newcommand*{\Acb}{\ensuremath{A_{\rm CB}}}
\newcommand*{\radsq}{\ensuremath{}}

\newcommand*{\BK}{BICEP/Keck}
\newcommand*{\litebird}{LiteBIRD}

\newcommand{\LCDM}{$\Lambda$CDM}

\newcommand*{\TB}{\ensuremath{T\!B}}
\newcommand*{\EB}{\ensuremath{E\!B}}
\newcommand*{\BB}{\ensuremath{B\!B}}

\newcommand*{\planck}{\textit{Planck}}

\newcommand*{\candl}{\texttt{candl}}

\graphicspath{{./}{figures/}}

\defcitealias{dunkley13}{D13}
\defcitealias{prince24}{P24}

\defcitealias{zebrowski25}{Z25}
\defcitealias{sayre20}{S20}

\defcitealias{namikawa24}{N24}
\defcitealias{lonappan25}{L25}

\shorttitle{Probing Anisotropic Cosmic Birefringence with Foreground-Marginalised SPT $B$-mode Likelihoods}
\shortauthors{L.~Balkenhol et al.}

\begin{document}

\journalinfo{The Open Journal of Astrophysics}

\title{Probing Anisotropic Cosmic Birefringence with\\Foreground-Marginalised SPT $B$-mode Likelihoods\vspace{-10ex}}

\author{L. Balkenhol\,\orcidlink{0000-0001-6899-1873}$^{1,\ast}$}
\email{$^\ast$lennart.balkenhol@iap.fr}

\author{A.~Coerver\,\orcidlink{0000-0002-2707-1672}$^{2}$}

\author{C.~L.~Reichardt\,\orcidlink{0000-0003-2226-9169}$^{3}$}

\author{J.~A.~Zebrowski\,\orcidlink{0000-0003-2375-0229}$^{4,5,6}$}

\affiliation{
$^1$Sorbonne Universit\'{e}, CNRS, UMR 7095, Institut d'Astrophysique de Paris, 98 bis bd Arago, 75014 Paris, France
}
\affiliation{$^2$Department of Physics, University of California, Berkeley, CA, 94720, USA}
\affiliation{$^3$School of Physics, University of Melbourne, Parkville, VIC 3010, Australia}
\affiliation{$^4$Kavli Institute for Cosmological Physics, University of Chicago, 5640 South Ellis Avenue, Chicago, IL, 60637, USA}
\affiliation{$^5$Department of Astronomy and Astrophysics, University of Chicago, 5640 South Ellis Avenue, Chicago, IL, 60637, USA}
\affiliation{$^6$Fermi National Accelerator Laboratory, MS209, P.O. Box 500, Batavia, IL, 60510, USA}

\begin{abstract}
In this work, we construct foreground-marginalised versions of the SPT-3G D1 and SPTpol cosmic microwave background (CMB) $B$-mode polarisation likelihoods.
The compression is performed using the CMB-lite framework and we use the resulting data sets to constrain anisotropic cosmic birefringence, parametrised by the amplitude of a scale-invariant anisotropic birefringence spectrum, $\Acb{}$.
Using the new SPT-3G data we report a $95\%$ upper limit on $\Acb{}$ of $ 1.2\times 10^{-4}\radsq{}$, which tightens to $0.53\times 10^{-4}\radsq{}$ when imposing a prior on the amplitude of gravitational lensing based on CMB lensing reconstruction analyses.
These are the tightest constraints on anisotropic birefringence from \BB{} power spectrum measurements to-date, demonstrating the constraining power of the South Pole Telescope.
The likelihoods used in this work are made publicly available via the \href{https://github.com/lbalkenhol/candl_data}{\texttt{candl\_data}} repository.
\end{abstract}

\maketitle



\section{Introduction}\label{sec:intro}

The $B$-mode polarisation of the cosmic microwave background (CMB) is rich in cosmological information.
Beyond potentially containing smoking gun evidence for inflation \citep{linde05, linde08, s4sciencebook, ellis23}, we may also find evidence for parity-violation \citep{komatsu22nat}, phase transitions \citep{greene24}, curvature perturbations \citep{ireland25} or primordial magnetic fields here \citep{Paoletti24, Khalife24}.
While the CMB $B$-mode polarisation is faint and contaminated by foregrounds, the signal from the gravitational lensing of $E$ modes has been confidently detected \citep{hanson13, naess14, bicep14a, bicep2keck15, keisler15, polarbear17, louis17, faundez20, sayre20, bicep2keck21, polarbear22, zebrowski25, louis25} and effort is being placed on better understanding, modelling, and mitigating the foregrounds \citep[for example][]{hoz20, vacher20, azzoni23, leloup23, Morshed24, wolz24, hertig24, vacher25, steier25}.




In this paper, we present foreground-marginalised $B$-mode likelihoods based on South Pole Telescope (SPT) observations.
We focus on the last two \BB{} power spectrum measurements released by the SPT collaboration \citep[][hereafter \citetalias{sayre20} and \citetalias{zebrowski25}, respectively]{sayre20, zebrowski25} and use the CMB-only likelihood framework developed by \citet{dunkley13}, which we refer to as \emph{CMB-lite,} to compress the multi-frequency data sets into CMB-only data vectors.
We use the compressed likelihoods to constrain anisotropic cosmic birefringence, i.e. a spatially-varying rotation of the polarization angle of CMB photons.
Generically, any pseudo-scalar field that couples to the electromagnetic field with a Chern-Simons term may cause such a rotation, making anisotropic birefringence a powerful way to look for new physics.
Axion-like particles fall into this category and are a particularly interesting candidate, as they arise in a plethora of different dark sector models and in the study of topological defects \citep{carroll89, carroll98, kosowsky96, Pospelov09, finelli09bire, fujita21, Kitajima22, Gonzalez23, Eskilt23, Galaverni23, pdgrpp24}.

This work is structured as follows.
In \S\ref{sec:lite} we provide background on the CMB-lite compression framework before applying it to the two SPT data sets.
In \S\ref{sec:cb} we report constraints on anisotropic cosmic birefringence.
We present our conclusions in \S\ref{sec:conclusion}.

\section{CMB-lite compression}
\label{sec:lite}

We briefly review the CMB-lite framework as introduced by \citet{dunkley13}.
As this technique has become commonplace, we keep this review short and refer the reader to the above work and subsequent applications for details \citep{calabrese13, planck15-11, choi20, prince24, balkenhol25, camphuis25, louis25}.

\begin{figure*}[ht!]
    \includegraphics[width=2.0\columnwidth]{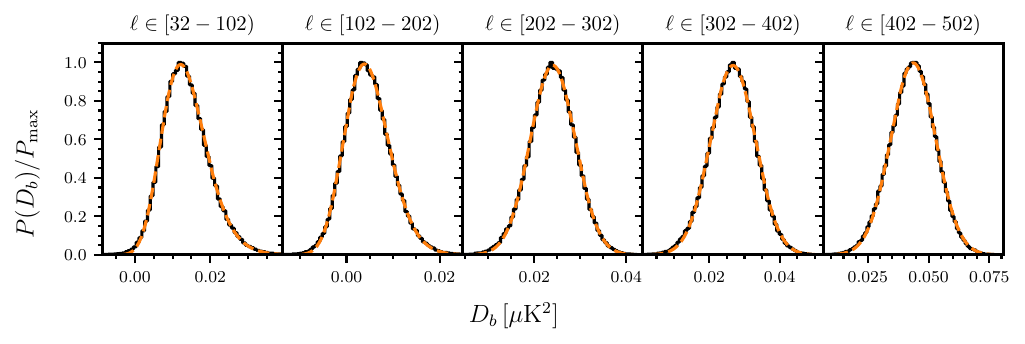}
    \caption{Histogram of MCMC samples of the common CMB power in the five SPT-3G D1 \BB{} band powers (black).
    The distribution is visibly non-Gaussian in the first two bins, which contain fewer modes.
    We fit offset log-normal distributions to the histograms (orange dashed lines), which serve as the basis for the compressed CMB-only likelihood.
    }
    \label{fig:bdp_sampling_3G}
\end{figure*}

A CMB power spectrum analysis usually yields a set of multi-frequency band powers, i.e. a binned power spectrum measurement $C^{\mu\nu,\mathrm{data}}_b$, contaminated by foreground emission and systematic effects.
Here $b$ indexes the band power bin, while $\mu$ and $\nu$ indicate the frequency channels of a given cross-frequency spectrum.
The CMB-lite framework models the multi-frequency band powers with CMB power $C_b^{\mathrm{CMB}}$ common to coinciding bins in the multi-frequency data vector as defined by the design matrix $A^{\mu\nu}$.
A parametric model $\theta$ for foreground emission and systematic effects then modifies the pure CMB signal.
Assuming for simplicity that the foreground contamination $C_b^{\mu\nu,\mathrm{FG}}(\theta)$ is additive (e.g. galactic dust) and systematic contamination $y^{\mu\nu}(\theta)$ multiplicative (e.g. calibration uncertainty) we have the following data model
\begin{align}
\label{eq:lite_model}
    C_b^{\mu\nu,\mathrm{model}}(C_b^{\mathrm{CMB}}, \theta) = y^{\mu\nu}(\theta) \bigg[ &A^{\mu\nu} C_b^{\mathrm{CMB}} \nonumber \\
    &+ C_b^{\mu\nu,\mathrm{FG}}(\theta) \bigg],
\end{align}
which is then used to construct a likelihood of the CMB signal, for example:
\begin{align}
\label{eq:recon_like}
    \mathcal{L} \propto &\left[ C_b^{\mu\nu,\mathrm{model}}(C_b^{\mathrm{CMB}}, \theta)-C^{\mu\nu,\mathrm{data}}_b\right]^{T} \Sigma^{-1} \nonumber \\
    &\left[ C_b^{\mu\nu,\mathrm{model}}(C_b^{\mathrm{CMB}}, \theta)-C^{\mu\nu,\mathrm{data}}_b\right],
\end{align}
where $\Sigma$ is the covariance matrix of the multi-frequency data vector.
Note that this likelihood does not depend on a cosmological model (beyond any model-dependence that may already present in the multi-frequency data).

The above likelihood can then be explored in different ways to obtain the values of the underlying CMB band powers and their covariance, for example through Markov chain Monte Carlo (MCMC) sampling \citep{dunkley13, calabrese13, planck15-11, choi20, prince24, louis25} or alternatively minimisation and a subsequent calculation of the Hessian at the best-fit point \citep{balkenhol25, camphuis25}.
This compressed CMB-only data set can then be used for cosmological inference.
The resulting lite likelihood comes with several advantages:
a speed-up of the likelihood through the reduction of the data vector, the facilitation of MCMC analysis through the elimination of nuisance parameters, and improved interpretability.

Note that the CMB-lite framework assumes that the CMB power is achromatic in the units of the underlying maps, i.e. the maps have been calibrated using the CMB black body spectrum into units of $K_{\mathrm{CMB}}$ \citep{fixsen96b}.
As such, cosmological models with a modified spectral energy distribution for the primordial signal of interest (e.g. Faraday rotation \citep{Paoletti24}) can not be constrained using compressed band powers as constructed above.
Instead, one would have to add an additional set of band powers to \cref{eq:lite_model} that obey the frequency scaling of the signal of interest.
This caveat does not apply to this work, as the model considered in \S\ref{sec:cb} does not modify the CMB frequency spectrum.

Below, we apply the CMB-lite framework separately to the SPT-3G D1 \BB{} \citepalias{zebrowski25} and SPTpol \BB{} releases \citepalias{sayre20}.
A joint reconstruction would necessitate quantifying the correlation of the two data sets, which is beyond the scope of this work.
The goal of this section is to create streamlined version of existing power spectrum measurements, rather than a new combined measurement.

\begin{figure*}[ht!]
    \includegraphics[width=2.0\columnwidth]{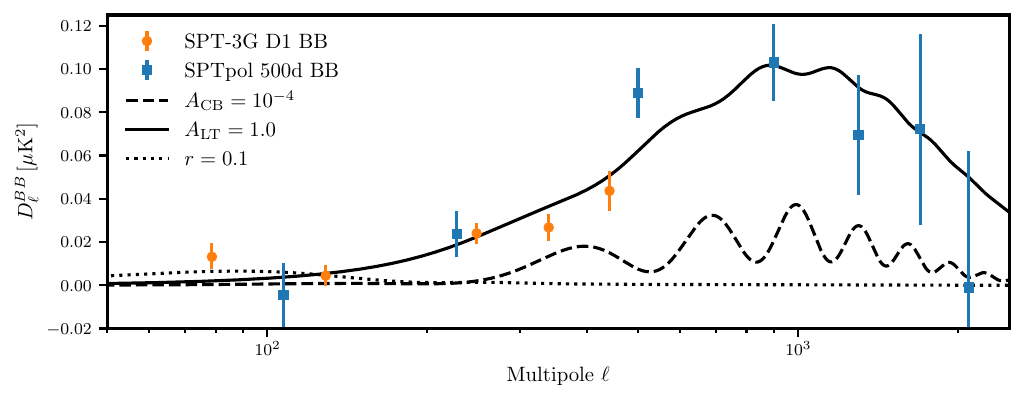}
    \caption{CMB-only band powers derived from SPT-3G D1 \BB{} data (orange, \citetalias{zebrowski25}) and SPTpol \BB{} data (blue, \citetalias{sayre20}).
    In contrast to the SPTpol data, the distribution of SPT-3G band powers on large scales is appreciably non-Gaussian;
    we indicate the median along with 16$\%$ and 84$\%$ intervals in this case.
    The black lines indicate the predicted power from the lensing of $E$ modes (solid line, $\Alens=1$), primordial gravitational waves (dotted line, $\rgw=0.1$), and anisotropic birefringence (dashed line, $\Acb=10^{-4}\radsq{}$).
    }
    \label{fig:BB_lite_bdp}
\end{figure*}

\subsection{SPT-3G D1 \BB{}}\label{sec:lite_3g}

We first apply the lite compression to the SPT-3G D1 \BB{} data release ~\citepalias{zebrowski25}{}.
The data set is based on two years of SPT-3G observations of a $1500\,\deg^2$ field in the southern sky at $95$, $150$, and $220\,\mathrm{GHz}$.
The analysis apodisation mask is chosen to match the \BK{} survey region to avoid foreground contamination as much as possible.
The resulting \BB{} band powers span the multipole range $32 \leq \ell < 502$ with five bins in each frequency cross-spectrum for a total of 30 multi-frequency band powers.
The multi-frequency likelihood has seven nuisance parameters that characterise the polarisation calibration of each frequency channel ($P_{\rm cal}^{95}$, $P_{\rm cal}^{150}$, $P_{\rm cal}^{220}$) and account for contamination due to galactic dust ($D_{\ell=80}^{\rm dust, 150}$, $\alpha^{\rm dust}$, $\beta^{\rm dust}$) as well as unresolved radio galaxies ($D_{\ell=500}^{\rm rg, 95}$).
We refer the reader to~\citetalias{zebrowski25}{} for details on the data model.

We perform the compression using the \candl{} implementation of the SPT-3G D1 \BB{} likelihood.\footnote{\url{https://github.com/SouthPoleTelescope/spt_candl_data}}
Since the window functions of coinciding band powers differ between the cross-frequency spectra, there exists no uniquely defined set of CMB-only band powers to reconstruct \citep{prince24}.
However, as demonstrated by~\citet{balkenhol25, camphuis25}, using an average window function with weights based on the band power covariance matrix leads to adequate performance.
We exploit the differentiability of the \candl{} likelihood and use No U-Turns (NUTS) sampling~\citep{hoffman11} to characterise the posterior of the five underlying CMB band powers plus the seven nuisance parameters.
In this way, we build up a total of $200,000$ samples, which we show in \cref{fig:bdp_sampling_3G}, with over $100,000$ effective samples in each band power bin.

As expected for data at such low multipole moments, the posterior distributions for the first two band powers are visibly non-Gaussian.
We hence follow the treatment of~\citet{prince24}{} and fit an offset log-normal distribution parametrised by $D_0,\,\mu,\,\sigma$ to the histogram of samples; for each bin $b$, $D_{b,0}$ is the best-fit offset that makes $\ln{D_b-D_{b,0}}$ normally distributed with mean $\mu_b$ and standard deviation $\sigma_b$.
We then transform the band power samples according to $\ln{D_b-D_0}$ and obtain their covariance $Q$, which we use to build a cosmological likelihood:
\begin{align}
    -2\mathcal{L} = &\left[\ln(X - D_0)-\mu\right]^T Q^{-1} \left[\ln(X - D_0)-\mu\right]\notag\\
    &-\sum_b \ln(X_b - D_{0,b}),
\end{align}
where $X$ are the binned model predictions.
Given the high number of MCMC samples, we forego a correction for sampling noise in the covariance matrix $Q$ \cite[e.g.][]{hartlap06}.
The derived CMB-only band powers are shown in~\cref{fig:BB_lite_bdp}.

The compressed cosmological likelihood performs well: repeating the analysis in \citetalias{zebrowski25} of fitting a simple template model for gravitational waves and lensing power with amplitudes $\rgw{}$ and $\Alens{}$ (see also \cref{eq:cosmo_model}), respectively, recovers the upper limit on $\rgw{}$ obtained by the multi-frequency likelihood to better than $1\,\%$. Similarly, the $\Alens{}$ posterior mean is shifted by $0.1\,\sigma$ and the standard deviation is the same to within $3\,\%$.
We compare the marginalised posterior distributions for $\rgw{}$ and $\Alens{}$ obtained from the multi-frequency and lite likelihoods in \cref{fig:lite_verification}.
Note that the two likelihoods have different functional forms: the multi-frequency likelihood uses the Hamimeche-Lewis prescription \citep{hamimeche08}, while the lite likelihood is lognormal.
Despite this, the parameter constraints match well.
We similarly see good performance of the compressed likelihood in the extended model space explored later in this work.
The compressed likelihood is made publicly available.\footnote{\url{https://github.com/lbalkenhol/candl_data}}

\subsection{SPTpol \BB{}}\label{sec:lite_pol}

We now turn our attention the SPTpol \BB{} data release ~\citepalias{sayre20}{}.
The data set is based on three seasons of SPTpol observations of a $500\,\deg^2$ field (contained entirely within the SPT-3G main survey field) at $95$ and $150\,\mathrm{GHz}$.
The resulting \BB{} band powers span the multipole range $52 < \ell < 2301$ with seven bins in each frequency cross-spectrum for a total of 21 multi-frequency band powers.
Fourteen nuisance parameters enter the likelihood to characterise the polarisation calibration of each frequency channel ($P_{\rm cal}^{95}$, $P_{\rm cal}^{150}$), the beam uncertainty ($B_\ell^{i}, i\in[1,7]$) and to account for contamination due to galactic dust ($D_{\ell=80}^{\rm dust, 150}$) along with three independent Poisson terms to account for unresolved radio galaxies ($D_{\ell=3000}^{\rm rg, 95\,\times\,95}$, $D_{\ell=3000}^{\rm rg, 95\,\times\,150}$, $D_{\ell=3000}^{\rm rg, 150\,\times\,150}$).
Details on the data model can be found in~\citetalias{sayre20}{}.

We begin by implementing the Fortran likelihood in Python in the \candl{} framework.
In doing so, we forego the marginalisation over the beam uncertainty, instead adding a fiducial contribution to the covariance matrix \citep{hou18}.
We find that this has a negligible impact on parameter constraints when repeating the template analysis of \citetalias{sayre20}.
The reimplemented likelihood recovers the $95\%$ upper limit of $\rgw{}$ of ~\citetalias{sayre20}{} to better than $1\,\%$.
The $\Alens{}$ posterior shrinks by $<2\%$ and its mean shifts by $0.05\,\sigma$.
The $\chi^2$ matches the original likelihood to $10\%$.
Note that as the fit quality of \LCDM{} to the SPTpol data is poor ~\citepalias{sayre20}{} any analysis looking closely at absolute $\chi^2$ values should be performed carefully and results cross-checked with the original Fortran likelihood.
The \candl{} reimplementation of the SPTpol likelihood is made publicly available in the same repository as the SPT-3G D1 \BB{} lite likelihood.

\begin{figure*}[ht!]
    \includegraphics[width=2.0\columnwidth]{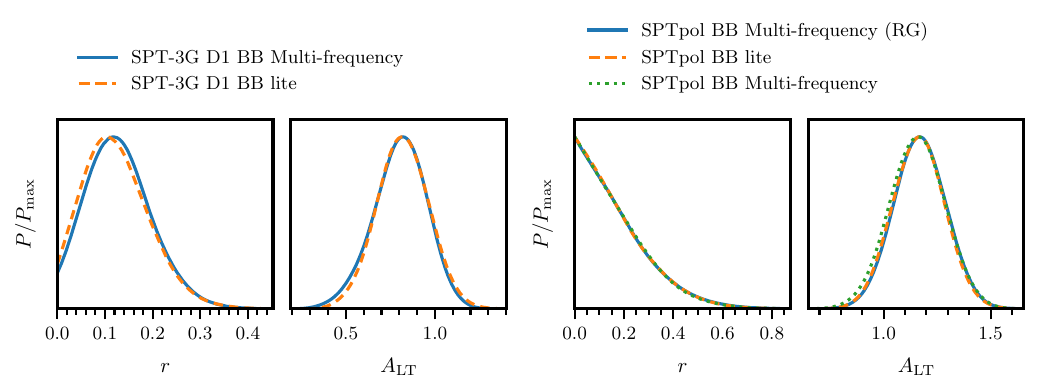}
    \caption{One-dimensional marginalised posterior distributions for the amplitude parameters $\rgw$ and $\Alens{}$ multiplying templates for gravitational waves and lensed $E$ modes, respectively, derived using multi-frequency and lite likelihoods.
    For both cases, the compressed lite likelihoods recover the constraints of the reference multi-frequency likelihoods within MC precision.\\
    \emph{Left:} SPT-3G D1 \BB{} multi-frequency analysis (blue solid line) compared to the lite likelihood (orange dashed line).
    \emph{Right:} SPTpol \BB{} multi-frequency likelihood using the radio-galaxy foreground model (blue solid line) used to build the lite likelihood (orange dashed line).
    For reference we indicate the constraints using the multi-frequency likelihood with the original foreground model (green dotted line).
    }
    \label{fig:lite_verification}
\end{figure*}

Before performing the lite compression we adjust the foreground model of the SPTpol \BB{} likelihood.
The independent Poisson terms of each cross-frequency spectrum in the multi-frequency likelihood lead to large degeneracies in the reconstruction.
This was already seen by~\citet{camphuis25} who note this for the case of temperature data.
Here, we replace the free Poisson terms by the radio galaxy model used in the SPT-3G likelihood, though fix the spectral index of the radio galaxies to the posterior peak $\alpha \sim -2.65$ found by \citetalias{sayre20}.\footnote{Note that as already pointed out by \citetalias{sayre20} this value is lower than what contemporary observations suggest \citep{mocanu13, gupta19, everett20}, though here we are only interested in producing a compressed likelihood that replicates the constraints of the multi-frequency likelihood as closely as possible.}
This change to the foreground model leads to minor shifts to parameter constraints in the above template analysis: the upper limit on $\rgw{}$ changes by less than $2\%$; the mean of the $\Alens{}$ posterior shifts down by $0.2\,\sigma$ while its width increases by $6\%$ as shown in \cref{fig:BB_lite_bdp}.

We now perform the lite compression of the \candl{} reimplementation of the SPTpol likelihood, using the updated foreground model.
We deal with window functions in the same way as for the SPT-3G likelihood.
The SPTpol bins are wider, meaning more modes enter each bin and as such the band power distributions can be well-approximated by a Gaussian, even for the first bin.
As such, we can perform the reconstruction in two ways, either by sampling the reconstruction likelihood, as we did for SPT-3G, or by simply minimising it and evaluating the Hessian, as introduced by~\citet{balkenhol25}{} and performed in~\citet{balkenhol25, camphuis25}.
As the likelihood is fast to evaluate, we perform both and find that the results match closely.
We proceed to build a Gaussian cosmological likelihood using the means and covariance of CMB band powers obtained from the MCMC samples.
For all seven CMB band power, we record over $130,000$ effective samples each and as such we again forego a correction for sampling noise in the covariance matrix.
The CMB-only band powers are shown in~\cref{fig:BB_lite_bdp}.

Repeating the same template analysis as before, we find that the posterior distributions derived from the compressed likelihood match the ones derived from the multi-frequency likelihood with the above radio-galaxy model well.
The lite likelihood recovers the corresponding upper limit on $\rgw{}$ to better than $2\,\%$. Similarly, the $\Alens{}$ posterior mean is shifted by less than $0.1\,\sigma$ and the standard deviation is unchanged to within $1\,\%$.
The marginalised posterior distributions for $\rgw{}$ and $\Alens{}$ obtained from the multi-frequency and lite likelihoods are shown in~\cref{fig:lite_verification}.
While the choice of $\alpha$ used during the reconstruction process based on results for \LCDM{} implies a conditioning of the compressed likelihood on the standard model, this is a small effect; we still see good performance of the compressed likelihood in the extended model space explored below.

\section{Anisotropic Birefringence}\label{sec:cb}

We now use the compressed likelihoods produced in the previous section to constrain anisotropic cosmic birefringence.
We begin by briefly reviewing the relevant background and explaining our methodology in \S\ref{sec:cb_background} before reporting results in \S\ref{sec:cb_results}.

\subsection{Background and Methodology}\label{sec:cb_background}

Cosmic birefringence leads to a rotation of the polarisation plane of CMB photons as they travel from the surface of last scattering to us today.
Isotropic birefringence describes a global rotation by the same angle across the sky; anisotropic birefringence describes the rotation by an angle varying across the sky.

While recent analyses of \EB{} data have shown some moderate evidence for isotropic birefringence \citep{Minami20, eskilt22, Eskilt22pl, Diego22, Eskilt23cg, Diego25}, it remains difficult to detect confidently as this hinges on the precise calibration of the polarisation angle of detectors.
Efforts to improve the calibration of ground-based experiments are being pursued \citep{ritaco24}.
Anisotropic birefringence on the other hand may be detected independently of a global rotation of the polarisation angle.
Limits have been derived from the analysis of \BB{} power spectra \citep{namikawa24, lonappan25}, as well as four-point correlation functions of the CMB \citep{gluscevic12, polarbear15, bicep2keck17, contreras17, namikawa20, bianchini20b, pasi20, Zagatti24}.

In this work, we look for the signature of anisotropic birefringence using SPT $B$-mode power spectrum measurements.
Importantly, the SPT-3G and SPTpol data sets are insensitive to a global rotation angle by construction.
The two analyses both employ an $EB$-leakage deprojection step during which any global rotation (be it cosmological or systematic) is undone.
Still, this procedure preserves the sensitivity of the data to an anisotropic rotation as it produces neither a \EB{} nor a \TB{} signal \citep{hongbo22}.

\citet{namikawa24} recently performed an exact calculation of the impact of anisotropic birefringence on the $B$-mode power spectrum, foregoing approximating the last scattering surface as thin.
We refer the reader to the above work for details of the calculation.
The end result is that $E$ modes are rotated into $B$ modes according to
\begin{align}
    C^{\rm BB,\, CB,\,aniso}_\ell = &4 \sum_{\ell\ L} p^+_{\ell L \ell'} \frac{(2\ell' + 1)(2L + 1)}{4\pi}\notag\\
    &\times \left(\begin{matrix}
\ell & L & \ell'\\
2 & 0 & -2
\end{matrix}\right)
\tilde{C}^{\rm EE}_{\ell' L},
\end{align}
where $\tilde{C}^{\rm EE}$ is the E-mode power spectrum, which is also distorted due to the birefringence effect, and $p^+_{\ell L \ell'} = [1+(-1)^{\ell+L+\ell'}]/2$.
We consider the rotation due to a scale-invariant birefringence spectrum, $C_L^{\alpha\alpha}$, which we define following the literature \citep[e.g.][]{namikawa24, bianchini20b} using the amplitude parameter \Acb{}:
\begin{align}
\label{eq:acb_def}
    \frac{L(L+1)}{2\pi} C_L^{\alpha\alpha} = \Acb{}\,\,\,[\mathrm{rad}^2].
\end{align}

Given the limited sensitivity of current data, we forego solving the full system of equations to calculate the expectation spectra for a given set of parameters.
Instead, we calculate a series of templates for the \planck{} 2018 best-fit cosmology and rescale their amplitude.
Our model is:
\begin{align}
\label{eq:cosmo_model}
        D^{\rm BB}_\ell = \rgw D^{\rm BB,\, r}_\ell + \Alens D^{\rm BB,\, lens}_\ell + \Acb D^{\rm BB,\, CB}_\ell,
\end{align}
where $\rgw{},\, \Alens{},\, \Acb{}$ are free parameters controlling the amplitudes of the pre-calculated templates $D^{\rm BB,\, r}_\ell,\, D^{\rm BB,\, lens}_\ell,\, D^{\rm BB,\, CB}_\ell$, which represent the signal from gravitational waves, lensed $E$ modes, and anisotropic birefringence, respectively.
We use the $D^{\rm BB,\, CB}_\ell$ template for the Planck 2018 best-fit cosmology by \citet{lonappan25} based on the modified CLASS version published by \citet{namikawa24} \citep{blas11}.
We recalculate the remaining templates ourselves using CAMB \citep{camb11}.
In \cref{fig:BB_lite_bdp}, we show all templates alongside the CMB-only SPT band powers constructed in the previous section.

We place uniform priors on the amplitude parameters $\rgw{},\Alens{},\Acb{}$, requiring them to be non-negative, unless otherwise explicitly stated.
We optionally apply a Gaussian prior of $\Alens{} \sim N(1.010, 0.016^2)$ based on the joint SPT-3G, ACTpol, and \planck{} lensing analysis of~\citet{qu25}, for which we use the short-hand \emph{\Alens{} APS prior}.
We perform MCMC analyses using Cobaya \citep{torrado21} and enforce a Gelman-Rubin convergence criterion of $R-1<0.02$.
When presenting constraints we either report mean values and 68\% confidence intervals or 95\% upper limits.
We carry out separate analyses of the SPT-3G and SPTpol likelihoods.
As mentioned in \S\ref{sec:lite}, a full joint analysis would require quantifying their correlation, which is beyond the scope of this work.

\subsection{Results}\label{sec:cb_results}

\begin{figure}[ht!]
    \includegraphics[width=1.0\columnwidth]{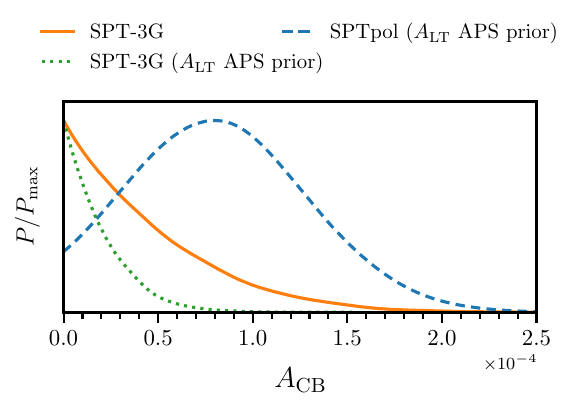}
    \caption{One-dimensional marginalised $\Acb{}$ posterior distributions derived using the SPT-3G D1 \BB{} lite likelihood without (solid orange) and with (dotted green) the $\Alens{}$ APS prior \citep{qu25}.
    We also show the SPTpol constraint with the same prior (dashed blue), which is consistent with results in the literature \citep{namikawa24, lonappan25}.
    The SPT-3G data place a tight $95\,\%$ upper limit on anisotropic birefringence of $\Acb{} < 1.2\,\times 10^{-4}\radsq{}$ without and $\Acb{} < 0.53\,\times 10^{-4} \radsq{}$ with the $\Alens{}$ APS prior.
    }
    \label{fig:cb_1d}
\end{figure}

We now report constraints placed by SPT data on the amplitude parameters $\rgw{},\Alens{},\Acb{}$.
Results for the SPTpol data have already been reported \citep{namikawa24, lonappan25} and we find consistent results with these works.\footnote{Using the same \planck{}-based prior on \Alens{} as \citet{namikawa24, lonappan25}, we report 95\% upper limits on $10^4 \times \Acb{}$ of $1.70$ $1.62$, $1.64$ for the multi-frequency likelihood with the original foreground model, the multi-frequency likelihood with the updated radio galaxy model, and the lite likelihood. These numbers match the literature results to within $15\%$. As noted in \S\ref{sec:lite_pol} the radio galaxy model leads to a small shift in \Alens{}, which due to the degeneracy of this parameter with \Acb{} is also reflected by a small change to the upper limit on \Acb{}.}
We therefore focus on the analysis of the new SPT-3G data here and only use the SPTpol results a comparison point when appropriate.
Using the SPT-3G D1 \BB{} lite likelihood we report
\begin{align}
\rgw &< 0.26,\notag\\
\Alens &= 0.69\,\pm\,0.18,\notag\\
\Acb &< 1.2\,\times 10^{-4}\radsq{}.
\end{align}
We show the marginalised posterior distribution for $\Acb{}$ in \cref{fig:cb_1d}.

\begin{figure}[ht!]
    \includegraphics[width=1.0\columnwidth]{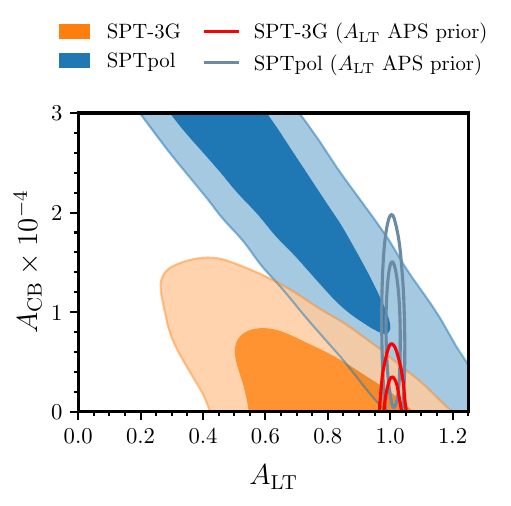}
    \caption{Posteriors in the $\Acb{}$-$\Alens{}$ plane derived from SPT-3G data (no \Alens{} prior: orange filled, \Alens{} APS prior: red line) and SPTpol data (no \Alens{} prior: blue filled, \Alens{} APS prior: grey line).
    Without an external prior on \Alens{} the two data sets constrain bands in the $\Acb{}$-$\Alens{}$ plane.
    The SPTpol data can support larger values in both parameters than the SPT-3G data, such that the SPTpol (SPT-3G) band intersects the $\Acb{}=0$ axis mostly above (below) $\Alens{}=1$.
    The tight $\Alens{}$ prior provided by the lensing reconstruction slices through the two bands such that in the case of SPT-3G we have a tight upper limit, whereas for SPTpol data we have a posterior that is mildly detached from $\Acb{}=0$.
    }
    \label{fig:alens_prior}
\end{figure}

In contrast to the SPTpol data, the SPT-3G data appear to be able to constrain $\Acb{}$ reasonably well without a prior on $\Alens$.
The correlation between these two parameters drops from $\rho(\Alens,\Acb)=-0.93$ for SPTpol to $\rho(\Alens,\Acb)=-0.59$ for SPT-3G.
However, this is due to the prior boundary at $\Acb{}=0$.
We expect $\Acb{}>0$ as on the relevant scales this parameter is proportional the amplitude of the auto power spectrum of the rotation field (see \cref{eq:acb_def} above and Eq. 16 in~\citet{namikawa24}).
However, it is interesting to explore the unphysical scenario of $\Acb<0$ in this case to better understand the constraints placed by the data;
when allowing for $\Acb<0$, the SPT-3G posterior widens to $10^{4}\sigma(\Acb{})=0.94\radsq{}$, such that the $95\%$ confidence region now spans $3.7\times10^{-4} \radsq{}$ in \Acb{} (compared to the upper limit of $1.2\times10^{-4} \radsq{}$ before).
Similarly, the $\Alens{}$ posterior widens by about $70\%$ to $\sigma(\Alens{})=0.31$ and the correlation coefficient now is $\rho(\Alens,\Acb)=-0.87$.
Though the SPT-3G data are binned more finely and hence have some sensitivity to the oscillations induced by the cosmic birefringence effect, their coverage in multipole space remains limited.
The SPTpol data on the other hand have a large coverage in multipole space, but the oscillations average out across the broad band power bins.
As a consequence, both data sets constrain bands in the $\Acb{}$-$\Alens{}$ plane as illustrated in \cref{fig:alens_prior}.
An external prior on \Alens{} can break this degeneracy.


\begin{figure}[hb!]
    \includegraphics[width=1.0\columnwidth]{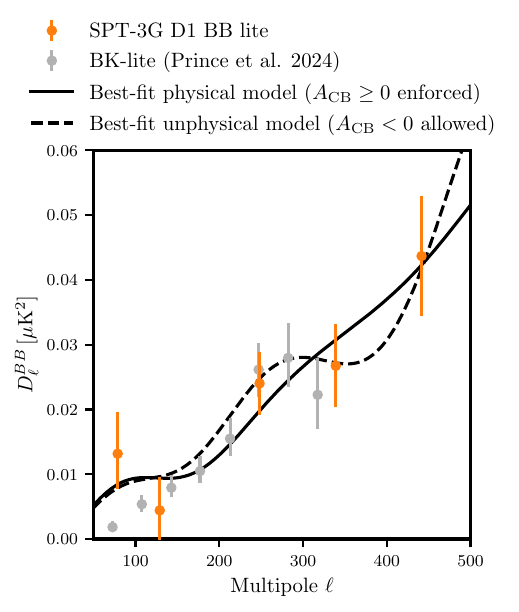}
    \caption{SPT-3G D1 \BB{} CMB-only band power (orange) along with the associated best-fit model spectra when allowing $\Acb$ to vary (solid black line).
    We also show the best-fit model spectrum when allowing $\Acb$ to take negative, unphysical values (dashed black line).
    The third bin of the SPT-3G data scatters high, while the fourth bin scatters low with respect to the baseline best-fit spectrum.
    By increasing $\Alens > 1$ the third bin can be better fit, while $\Acb<0$ introduces a dip that fits the fourth band power better.
    The two effects average out across the final band power bin.
    In grey, we show the BK-lite band powers derived by \citet{prince24}.
    While these data appear to follow a similar trend across $\ell=250-350$, the SPT and \BK{} analyses share the same analysis mask and are therefore correlated.
    Given that the noise in the SPT and \BK{} data is uncorrelated, this suggest that the trend seen by the two experiments may be due to a sample variance fluctuation in the shared survey field.
    }
    \label{fig:cb_bf}
\end{figure}

We now repeat the above analysis, this time imposing the $\Alens{}$ APS prior and enforcing $\Acb\geq0$ once again.
We show the associated one-dimensional $\Acb{}$ posterior in \cref{fig:cb_1d} and the two-dimensional $\Acb{}$-$\Alens{}$ one in \cref{fig:alens_prior}.
For SPT-3G, we report:
\begin{align}
\rgw &< 0.26\notag\\
\Alens &= 1.007\,\pm\,0.016\notag\\
\Acb &< 0.53\,\times 10^{-4}\radsq{}.
\end{align}
This is the tightest constraint on anisotropic birefringence based on \BB{} power spectrum measurements to-date.
The $\Acb{}$ limit above is approximately three times tighter than the corresponding SPTpol result (we find $\Acb < 1.6\,\times 10^{-4} \radsq{}$), though the $\Acb$ posterior peaks at $\Acb>0$ in this case.
Moreover, the SPT-3G limit is tighter than the limit placed by the combination of \BK{}, Polarbear, and ACT data reported by \citet{lonappan25} of $\Acb < 0.85\,\times 10^{-4} \radsq{}$,\footnote{Note that this limit is derived using a \planck{}-based prior on \Alens{}. However, replacing the APS prior with this one for SPT-3G only loosens the upper limit to $\Acb < 0.55 \,\times 10^{-4}\radsq{}$.} though note that the SPT-3G and \BK{} data are correlated.
The above constraint is impressive as it only uses $B$-mode information at the power spectrum level.
Still, tighter constraints can be derived by using the full polarisation field: \citet{bianchini20b} report $\Acb < 0.1\,\times 10^{-4} \radsq{}$ in a dedicated anisotropic birefringence analysis of SPTpol data.
Given the results above, we conclude that we see no evidence for anisotropic cosmic birefringence in the SPT \BB{} data.



To better understand the tight limit placed by SPT-3G data on cosmic birefringence, we return to the previous analysis allowing for negative, unphysical \Acb{} values (without the $\Alens{}$ APS prior).
We find $\Acb = (-1.13\,\pm\,0.94)\,\times 10^{-4} \radsq{}$, which is compatible with zero at $1.2\,\sigma$.
Still, $89\%$ of the posterior mass lies at $\Acb<0$; this region is not explored in the baseline analysis enforcing non-negative $\Acb$ and hence the posterior is forced against the boundary.\footnote{Readers may be familiar with a similar effect when looking at constraints on the sum of neutrino masses derived from CMB data (and in particular involving \planck{} temperature data).}
We compare the best-fit model spectrum when allowing for $\Acb<0$ to the one of the baseline analysis in \cref{fig:cb_bf}.
As the plot reveals, the third bin of the SPT-3G data scatters high, while the fourth bin scatters low with respect to the best-fit baseline model ($\Acb\geq 0$) spectrum.
The upwards scatter of the third bin can be well fit by increasing $\Alens{}$ slightly.
In isolation this would lead to an excess in power in the fourth and fifth bins.
However, the oscillatory shape of the birefringence template is important here, as it features a peak between the last two band power bins ($300\lesssim \ell \lesssim 500$).
By subtracting the template, the low fourth band power is better fit.
At the same time, the increase in $\Alens$ balances out this reduction in power for the last band power bin.
The oscillations introduced by the the cosmic birefringence template average out around $\ell\sim450$, which coincides with the effective centre of the final bin.
Therefore, with $\Alens{}>1$ and $\Acb<0$ the model is able to predict more power in the third bin and less power in the fourth bin, without spoiling the fit to the other band powers.
Conversely, $\Acb>0$ leads to the opposite effect and is therefore disfavoured by the data, which leads to the tight limits on $\Acb$ we found above.
We stress that the scatter of the SPT-3G band powers is entirely within statistical expectations; the data are well described by a model with only $\rgw$ and $\Alens{}$ \citepalias{zebrowski25} and even when allowing for $\Acb{}<0$ the constraint remains compatible with zero at $1.2\,\sigma$.

\section{Conclusions}\label{sec:conclusion}

In this work, we have presented foreground-marginalised, CMB-only likelihoods based on the SPT-3G D1 \BB{} and the SPTpol \BB{} data sets.
The compression of the multi-frequency data is done using the lite framework and, for the first time, accelerated by gradient-based sampling.
We make these compressed likelihoods publicly available, along with a python reimplementation of the SPTpol multi-frequency likelihood.

We use the compressed likelihoods to explore constraints on anisotropic cosmic birefringence, finding no evidence for such a rotation in the SPT data.
Using the new SPT-3G data we derive upper limits of $\Acb{} < 1.2 \,\times 10^{-4}\radsq{}$ without and $\Acb{} < 0.53\,\times 10^{-4} \radsq{}$ with an external prior on the amplitude of gravitational lensing.
The latter is the tightest constraint on anisotropic birefringence derived from \BB{} power spectrum measurements to date, demonstrating the constraining power of SPT-3G data.
Future CMB polarisation data from SPT, as well as \BK{}, the Simons Observatory and the \litebird{} mission \citep{lb25} will be even more sensitive to anisotropic birefringence.

\begin{acknowledgments}

L.B. is deeply grateful to F. Bianchini, S. Galli, M. Lembo, and T. Namikawa for helpful discussions and to T. Namikawa, A. I. Lonappan, B. Keating, and K. Arnold for making their work public.
The authors would also like to thank the South Pole Telescope collaboration for their support.

This work uses JAX \citep{jax18} and the scientific python stack \citep{jones01, hunter07, vanDerWalt11}.
This project has received funding from the European Research Council (ERC) under the European Union’s Horizon 2020 research and innovation programme (grant agreement No 101001897).
This work has received funding from the Centre National d’Etudes Spatiales and has made use of the Infinity Cluster hosted by the Institut d’Astrophysique de Paris. 
A.C. was supported by the National Science Foundation Graduate Research Fellowship Program under grant No. DGE 1752814 and by the National Science Foundation through the award OPP1852617.
C.R. acknowledges support from the Australian Research Council’s Discovery Project scheme (No. DP210102386).
Support for this work for J.Z. was provided by NASA through the NASA Hubble Fellowship grant HF2-51500 awarded by the Space Telescope Science Institute, which is operated by the Association of Universities for Research in Astronomy, Inc., for NASA, under contract NAS5-26555.
The South Pole Telescope program is supported by the National Science Foundation (NSF) through awards OPP-1852617 and OPP-2332483.
Partial support is also provided by the Kavli Institute of Cosmological Physics at the University of Chicago.

\end{acknowledgments}


\bibliographystyle{aa}
\typeout{}
\bibliography{BBlite}


\end{document}